\documentclass[prd,onecolumn]{revtex4}
\usepackage{dcolumn}
\usepackage{multirow}
\usepackage{graphicx}
\usepackage{amssymb}
\usepackage{bm}
\usepackage{hyperref}
\usepackage{epstopdf}
\usepackage{color}
\usepackage{mathrsfs}
\usepackage{amsmath,amssymb,amsthm}
\usepackage{rotating}
\usepackage{sverb, longtable}
\usepackage{subfigure}

\usepackage[graphicx]{realboxes}
\usepackage{adjustbox}
\begin{document}
\title{4D Gauss-Bonnet gravity: cosmological constraints, $H_0$ tension and large scale structure}
\author{Deng Wang}
\email{cstar@nao.cas.cn}
\affiliation{National Astronomical Observatories, Chinese Academy of Sciences, Beijing, 100012, China}

\author{David Mota}
\affiliation{Institute of Theoretical Astrophysics, University of Oslo, P.O. Box 1029 Blindern, N-0315
	Oslo, Norway}
\begin{abstract}
We perform correct and reasonable cosmological constraints on the newly proposed 4D Gauss-Bonnet gravity. Using the joint constraint from cosmic microwave background, baryon acoustic oscillations, Type Ia supernovae, cosmic chronometers and redshift space distortions, we obtain, so far,  the strongest constraint $\tilde{\alpha}=(1.2\pm5.2)\times 10^{-17}$, namely $\alpha=(2.69\pm11.67)\times10^{48}$ eV$^{-2}$, among various observational limitations from different information channels, which is tighter than previous bound from the speed of gravitational wave by at least one order of magnitude. We find that our bound is well supported by the observations of temperature and lensing potential power spectra of cosmic microwave background from the Planck-2018 final release. Very interestingly, the large $H_0$ tension between the local measurement from the Hubble Space Telescope and global derivation from the Planck-2018 final data under the assumption of $\Lambda$CDM can be greatly resolved from $4.4\sigma$ to $1.94\sigma$ level in the 4D Gauss-Bonnet gravity. 
In theory, we find that this model can partly relieve the coincidence problem and the rescaling Gauss-Bonnet term, which needs the help of the cosmological constant to explain current cosmic acceleration, is unable to serve as dark energy alone.

\end{abstract}
\maketitle

\section{Introduction}
Up to now, general relativity (GR) is believed to be the most successful gravity theory to describe the physical and cosmological phenomena over a large range of energy from large scales to small scales \cite{Einstein:1916vd}. However, it is not a perfect theory based on observational limitations and theoretical considerations. Specifically, two dark clouds of modern physics, dark matter \cite{Young:2016ala} and dark energy \cite{Weinberg:1988cp,Riess:1998cb,Perlmutter:1998np}, can not be well explained in the framework of GR plus $\Lambda$-cold dark matter ($\Lambda$CDM), where $\Lambda$ is the cosmological constant. Meanwhile, the quantum version of Einstein's gravity, which meets at least the re-normalization problem, can not be reasonably constructed \cite{tHooft:1974toh,Deser:1974cy,Deser:1974cz,Goroff:1985sz,Goroff:1985th,vandeVen:1991gw}. This implies that the underlying gravity theory governing the gravitational dynamics of the universe may not be GR and could be an alternative gravitational scenario, which can help understand the dark sector better at least.              

The Lovelock's theorem \cite{Lanczos:1938sf,Lovelock:1971yv,Lovelock:1972vz} indicates that GR is the sole gravity theory under the following four assumptions: (i) metricity; (ii) diffeomorphism invariance; (iii) equations of motion must be second order; (iv) spacetime must be $3+1$ dimension. In general, one can construct a new gravitational theory by relaxing these assumptions. As an elegant variant or modification of GR, Gauss-Bonnet gravity \cite{Zumino:1985dp} satisfies the former three assumptions but just lives in high dimensions, and its corresponding action in $D$ dimensions is written as 
\begin{equation}
S = \int d^Dx \sqrt{-g}\left(\frac{R-2\Lambda}{16\pi G}+\alpha \mathcal{G}\right)+S_m,       \label{1}
\end{equation}  
where $R$ is the Ricci Scalar, $G$ is the Newton gravitational constant, $S_m$ is the action of matter fields, $\alpha$ is a constant and $\mathcal{G}$ is the so-called Gauss-Bonnet invariant read as   
\begin{equation}
\mathcal{G} = R^{\mu\nu\alpha\beta}R_{\mu\nu\alpha\beta}-4R^{\mu\nu}R_{\mu\nu}+R^2.  \label{2}
\end{equation}  

As we know, Gauss-Bonnet invariant is the unique form constructed from
quadratic products of Riemann tensor that does not introduce any terms with more than two derivatives into the gravitational field equations \cite{Lovelock:1971yv}. In 4-dimensional spacetime, since it is a total derivative, it can not make contributions to gravitational dynamics. Nonetheless, this does not means that Gauss-Bonnet invariant is completely useless in 4 dimensions. Because of its topological properties, it can be used to classify topologies when doing the path-integral quantization of gravity. Recently, in light of the fact that the contribution of Gauss-Bonnet term to equations of motion is proportional to $(D-4)$ \cite{Mardones:1990qc,Torii:2008ru}, a new 4-dimensional Gauss-Bonnet theory (hereafter GB) is proposed in Ref.\cite{Glavan:2019inb} by rescaling the coupling constant $\alpha\rightarrow\alpha/(D-4)$ in order to produce nontrivial contributions to gravitational dynamics. Some interesting results have been obtained based on this new GB model \cite{Haghani:2020ynl,Narain:2020qhh,Gurses:2020ofy,Aoki:2020iwm,Clifton:2020xhc,Feng:2020duo,Garcia-Aspeitia:2020uwq}. In particular, several groups have placed observational constraints on the free parameter $\alpha$ of this GB model. Clifton {\it et al.} \cite{Clifton:2020xhc} find that the inflation of the early universe rules out almost all the negative values of $\alpha$ except extremely small negative ones, and that the system of binary black holes gives the constraint $0\lesssim\alpha\lesssim10^{21}\mathrm{eV}^{-2}$. However, Feng {\it et al.} \cite{Feng:2020duo} point out that the GB model with a bare vanishing $\Lambda$ has been ruled out by current cosmological and gravitational wave observations. They give $-3.49\times10^{50}\mathrm{eV}^{-2}\leqslant\alpha\leqslant1.49\times10^{51}\mathrm{eV}^{-2}$ according to the tight constraint on the speed of gravitational waves from the first detection of an electromagnetic
counterpart (GRB 170817A) to the gravitational wave signal (GW170817). Specially, Garc\'{a}-Aspeitia {\it et al.} \cite{Garcia-Aspeitia:2020uwq} give the estimated value $\alpha\approx1.604^{+0.017}_{-0.018}\times10^{62}\mathrm{eV}^{-2}$ by using the joint constraint of five cosmological probes at the background level, but their result is very inconsistent with the constraint from the speed of gravitational waves. Therefore, we are aiming at giving a correct and reasonable constraint on the typical parameter $\alpha$ of the GB model in light of current cosmological observations. Our constraint will be compatible with observational limitations from Refs.\cite{Clifton:2020xhc,Feng:2020duo}.  
By using a data combination of cosmic microwave background (CMB), baryon acoustic oscillations (BAO), Type Ia supernovae (SNe Ia), cosmic chronometers and redshift space distortions (RSD), we obtain, so far, the strongest constraint $\alpha=(2.69\pm11.67)\times10^{48}$ eV$^{-2}$ among various observational limitations from different information channels. 

This study is outlined in the following manner. In Section II, we display the cosmological equations of the GB model. In section III, we discuss the coincidence problem and investigate the behaviors of the effective equation of state of dark energy in the GB model. In Section IV, we describe the observational datasets and implement the cosmological constraints. In Section IV, we investigate the behaviors of large scale structure of the GB model. The discussions and conclusions are presented in the final section.   

\section{Cosmological equations}
In 4-dimensional spacetime, this new GB model makes the Gauss-Bonnet term produce a nontrivial contribution to the background evolution of the universe. By inserting Eq.(2) into Eq.(1) and varying the action Eq.(\ref{1}), the D-dimensional modified Einstein equation is shown as 
\begin{equation}
R_{\mu\nu}-\frac{1}{2}g_{\mu\nu}+\Lambda g_{\mu\nu}+\frac{\alpha}{D-4}\left(4R_{\mu\alpha\beta\sigma}R^{\;\,\alpha\beta\sigma}_\nu-8R_{\mu\alpha\nu\beta}R^{\alpha\beta}-8R_{\mu\alpha}R^{\;\,\alpha}_{\nu}+4RR_{\mu\nu}-g_{\mu\nu}\mathcal{G}\right)=8\pi GT_{\mu\nu}    ,     \label{3}
\end{equation} 
where $g_{\mu\nu}$ and $T_{\mu\nu}$ are the spacetime metric and energy-momentum tensor of matter fields, respectively. In a spatially flat 4-dimensional ($D\rightarrow4$) Friedmann-Robertson-Walker (FRW) universe (see also \cite{Glavan:2019inb,Haghani:2020ynl,Narain:2020qhh}) Friedmann equation of the GB model is expressed as   
\begin{equation}
H^2+2\alpha H^4 =\frac{8\pi G}{3}(\rho_b+\rho_{cdm}+\rho_r)+\frac{\Lambda}{3}, \label{4}
\end{equation}
where $H$ denotes the Hubble parameter and $\rho_b$, $\rho_{cdm}$ and $\rho_r$ are the energy densities of baryons, CDM and radiation components, respectively. Assuming a perfect fluid with the equation of state (EoS) $p_i=\omega_i\rho_i$, where $i=b$, $cdm$ and $r$, the energy conservation equation of the GB model can be written as      
\begin{equation}
\dot{\rho}_i+3H(p_i+\rho_i)=0, \label{5}
\end{equation}
where the dot denotes the derivative with respect to the cosmic time $t$.
Notice that for the independent component $\Lambda$, the EoS reads as $p_\Lambda=-\rho_\Lambda$. In order to constrain the GB model at background level, it is convenient to write its dimensionless Hubble parameter $E(z)$ as follows
\begin{equation}
E^2(z)+2\alpha H_0^2E^4(z)=(\Omega_{b0}+\Omega_{cdm0})(1+z)^3+\Omega_{r0}(1+z)^4+\Omega_{\Lambda0}, \label{6}
\end{equation}   
where $E(z)\equiv H(z)/H_0$, $H_0$ is the Hubble constant, and $\Omega_{b0}$, $\Omega_{cdm0}$, $\Omega_{r0}$ and $\Omega_{\Lambda0}$ are the present-day energy density ratio of baryons, CDM, radiation and cosmological constant, respectively. 
Note that the present-day density ratio of matter $\Omega_{m0}=\Omega_{b0}+\Omega_{cdm0}$. 
To carry out constraints more conveniently, we define a dimensionless parameter $\tilde{\alpha}\equiv2\alpha H_0^2$ here. Solving the above equation with respect to $E(z)$ at a given redshift $z$, one can easily obtain
\begin{equation}
E(z)=\left[\frac{\sqrt{X(z)}-1}{2\tilde{\alpha}}\right]^\frac{1}{2}, \label{7}
\end{equation}
where 
\begin{equation}
X(z)=1+4\tilde{\alpha}\left[\Omega_{m0}(1+z)^3+\Omega_{r0}(1+z)^4+\Omega_{\Lambda0}\right]. \label{8}
\end{equation}
Using the condition $E(0)=1$, one can obtain 
\begin{equation}
1+\tilde{\alpha}=\Omega_{m0}+\Omega_{r0}+\Omega_{\Lambda0}. \label{9}
\end{equation}
It is worth noting that this equation indicates that $\tilde{\alpha}$ is a very small constant close to zero.
In the following context, this relation can help determine the range of the model parameter $\tilde{\alpha}$.

\begin{figure}[htbp]
	\centering
	\includegraphics[scale=0.6]{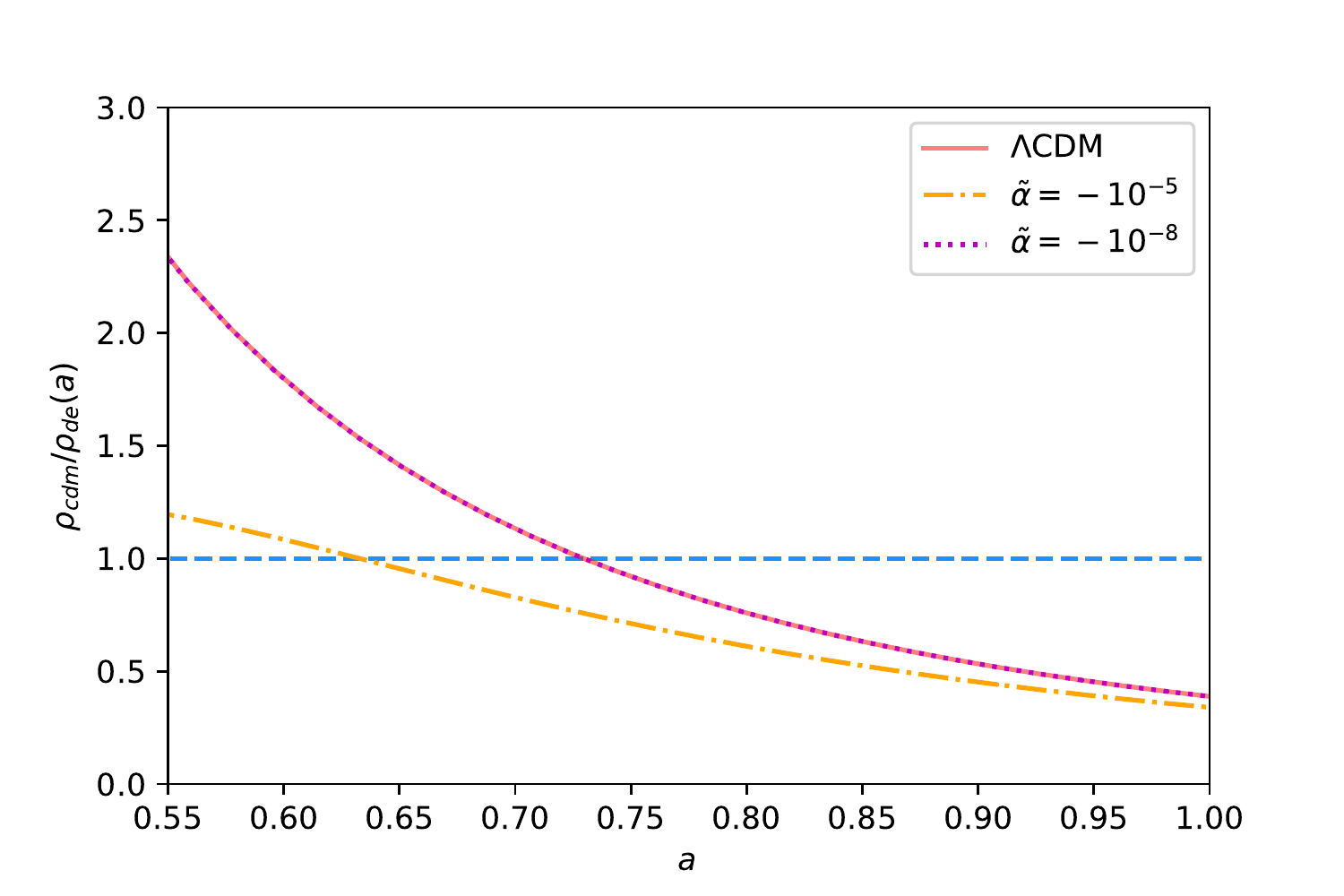}
	\caption{The density ratio between dark matter and dark energy is shown as a function of scale factor $a$. The solid (red), dash-dotted (orange) and dotted (magenta) lines denote $\tilde{\alpha}=0$ ($\Lambda$CDM), $-10^{-5}$ and $-10^{-8}$, respectively. The horizontal dashed (blue) line represents dark matter and dark energy share the same energy density. }\label{f1}
\end{figure}

\begin{figure}[htbp]
	\centering
	\includegraphics[scale=0.58]{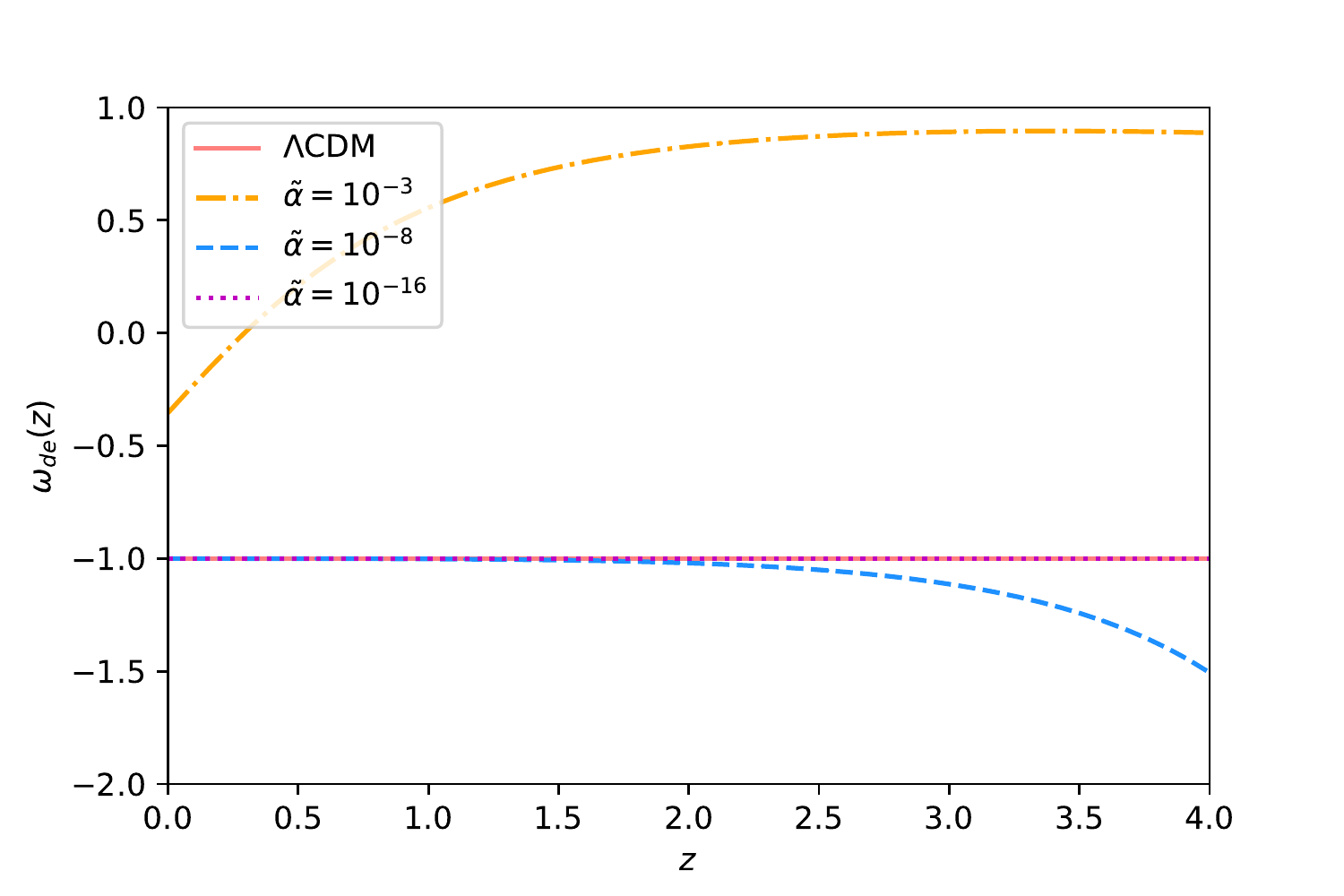}
	\includegraphics[scale=0.58]{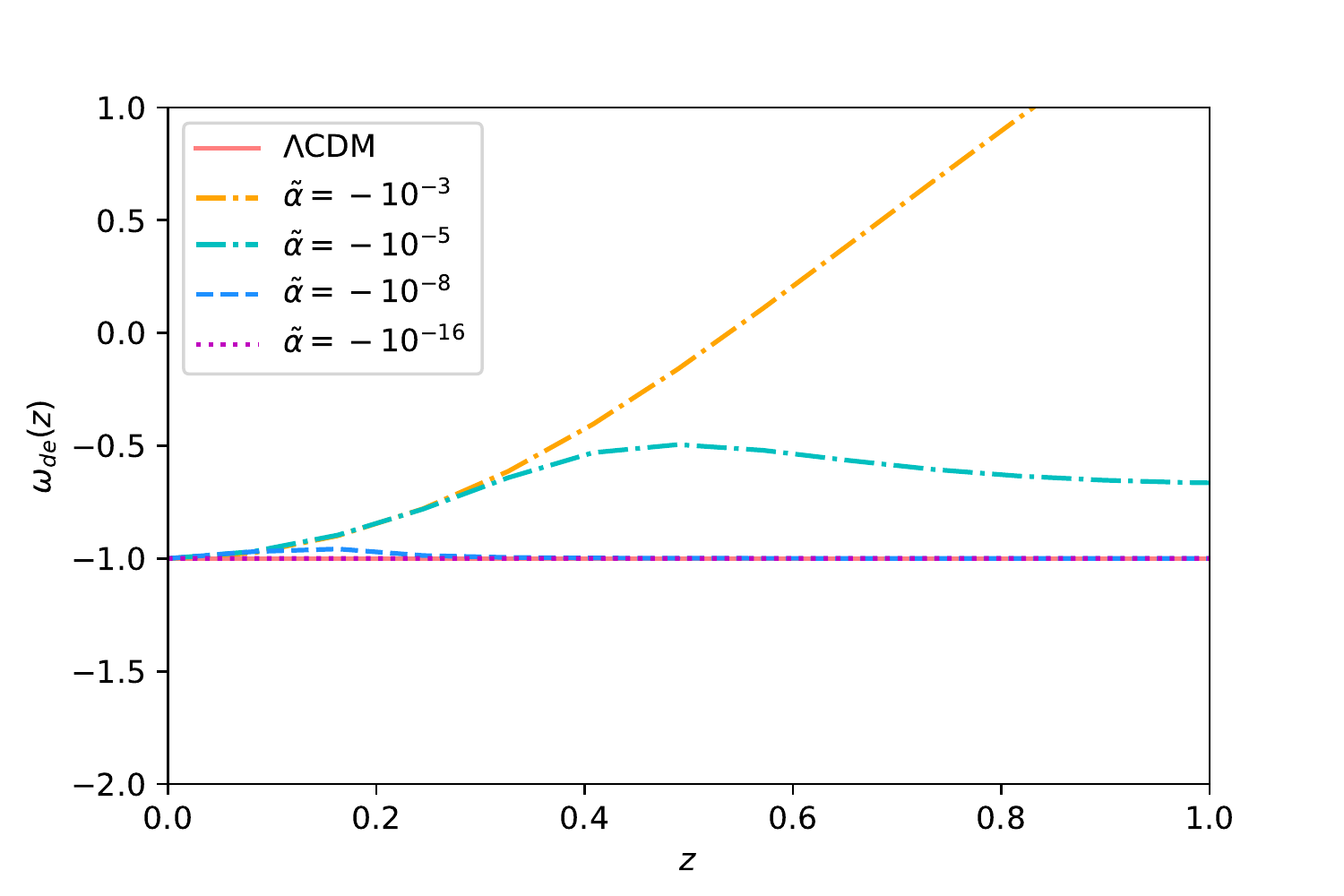}
	\caption{The effective EoS of dark energy $\omega_{de}(z)$ is shown as a function of redshift $z$. The solid (red) lines denote the $\Lambda$CDM model. {\it Left}: The dash-dotted (orange), dashed (blue) and dotted (magenta) lines are $\tilde{\alpha}=-10^{-3}$, $10^{-8}$ and $10^{-16}$ in the GB model, respectively. {\it Right}: The dash-dotted (orange), dash-dotted (cyan), dashed (blue) and dotted (magenta) lines are $\tilde{\alpha}=-10^{-3}$, $10^{-5}$, $10^{-8}$ and $10^{-16}$ in the GB model, respectively. }\label{f2}
\end{figure}

\begin{figure}[htbp]
	\centering
	\includegraphics[scale=0.7]{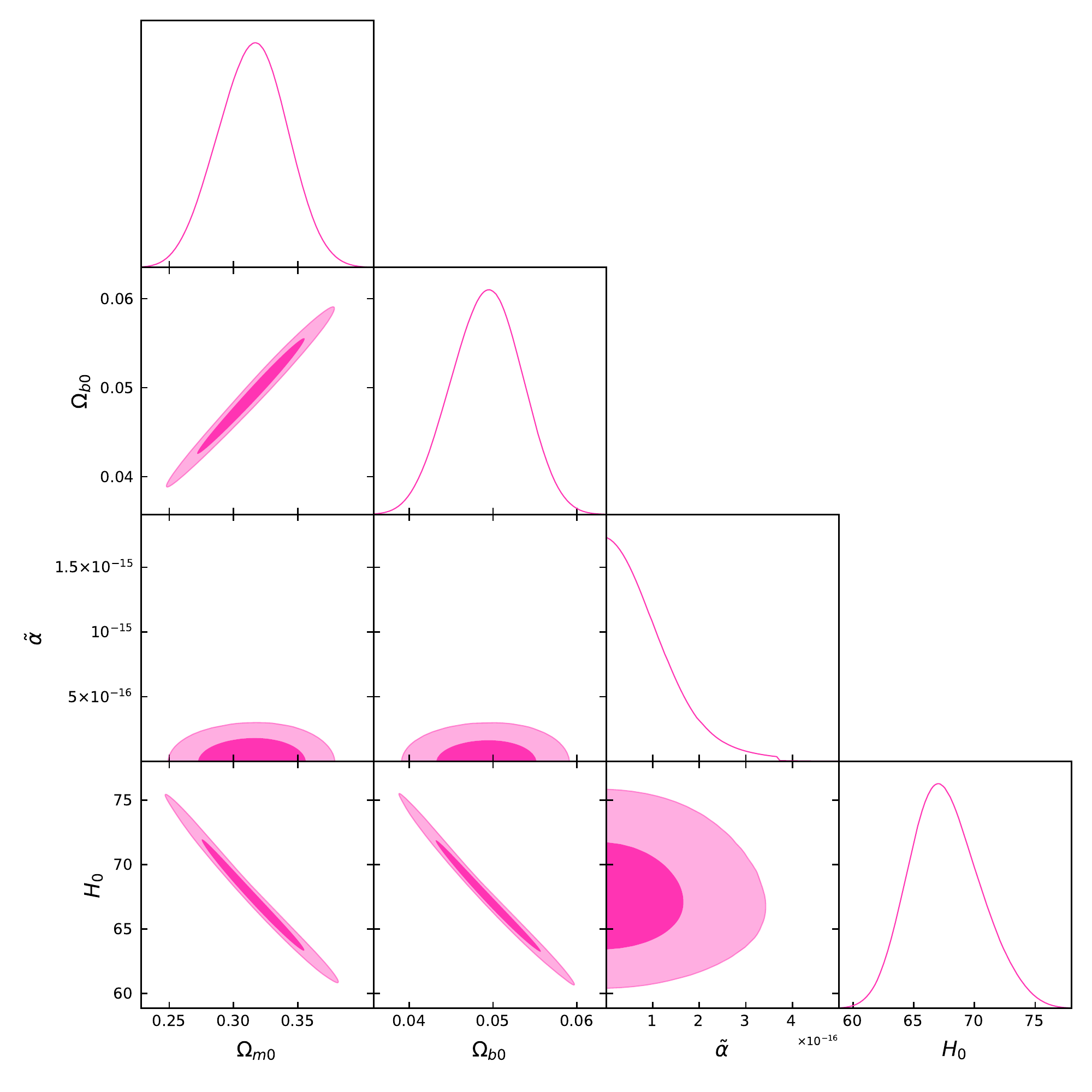}
	\caption{The marginalized constraint on the PGB model is shown by using the ``C'' dataset. }\label{f3}
\end{figure}

\begin{figure}[htbp]
	\centering
	\includegraphics[scale=0.7]{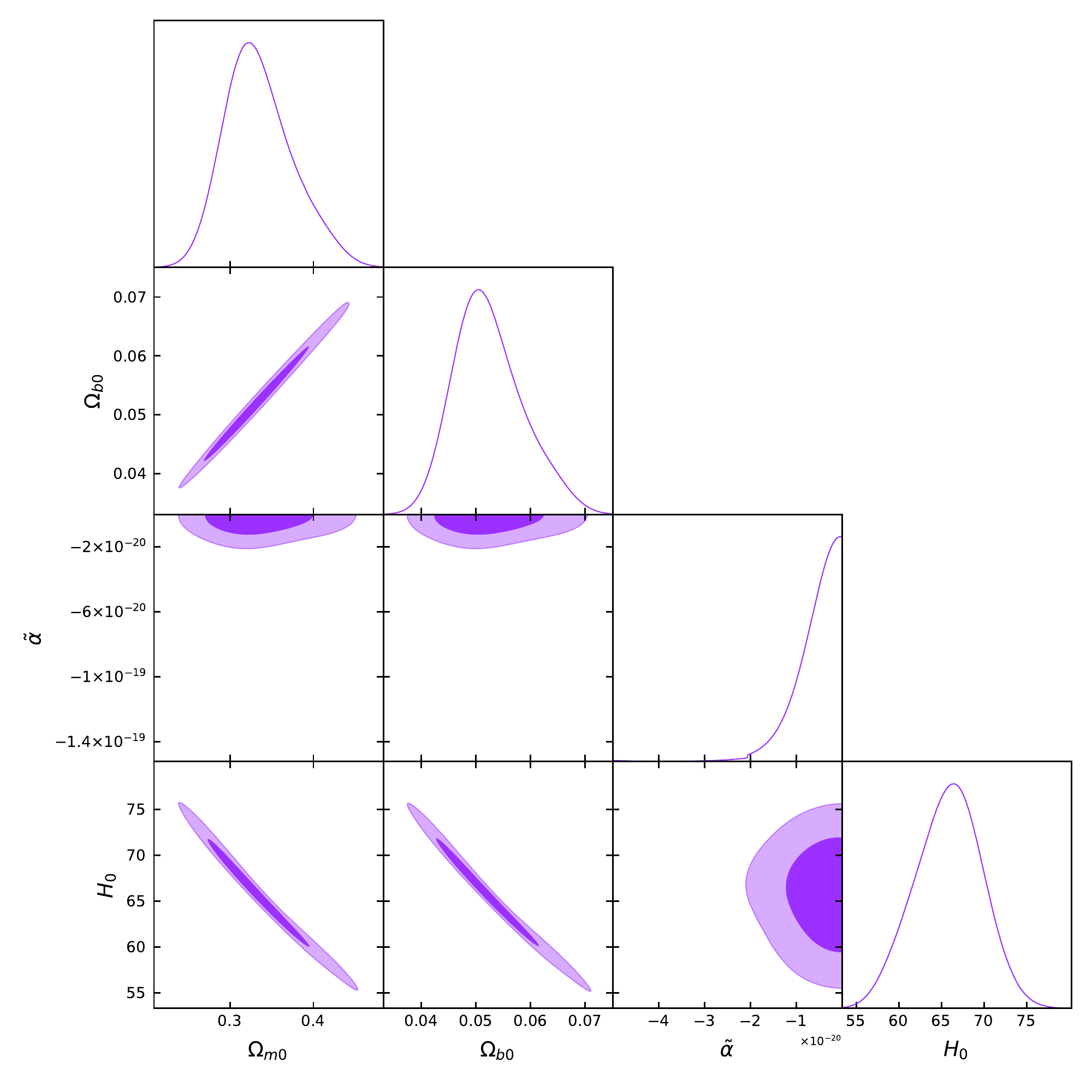}
	\caption{The marginalized constraint on the NGB model is shown by using the ``C'' dataset.}\label{f4}
\end{figure}

\begin{figure}[htbp]
	\centering
	\includegraphics[scale=0.45]{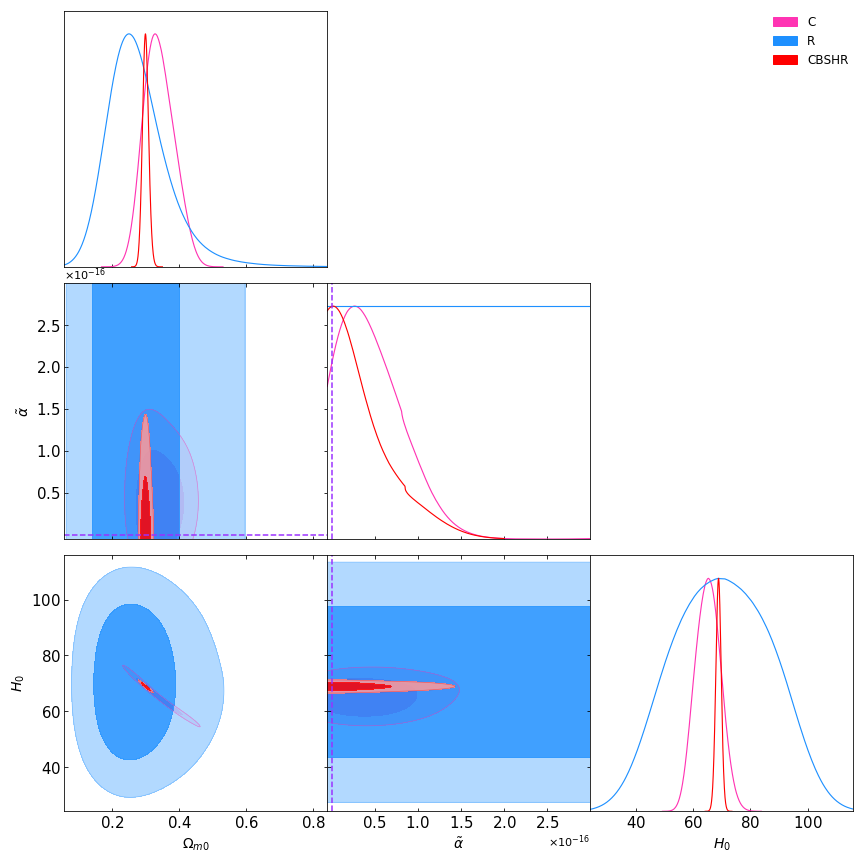}
	\caption{The marginalized constraints on the GB model are shown by using the ``C'', ``R'' and ``CBSHR'' datasets, respectively. The dashed magenta line denotes $\tilde{\alpha}=0$, i.e., the $\Lambda$CDM model.}\label{f5}
\end{figure}

\begin{figure}[htbp]
	\centering
	\includegraphics[scale=0.6]{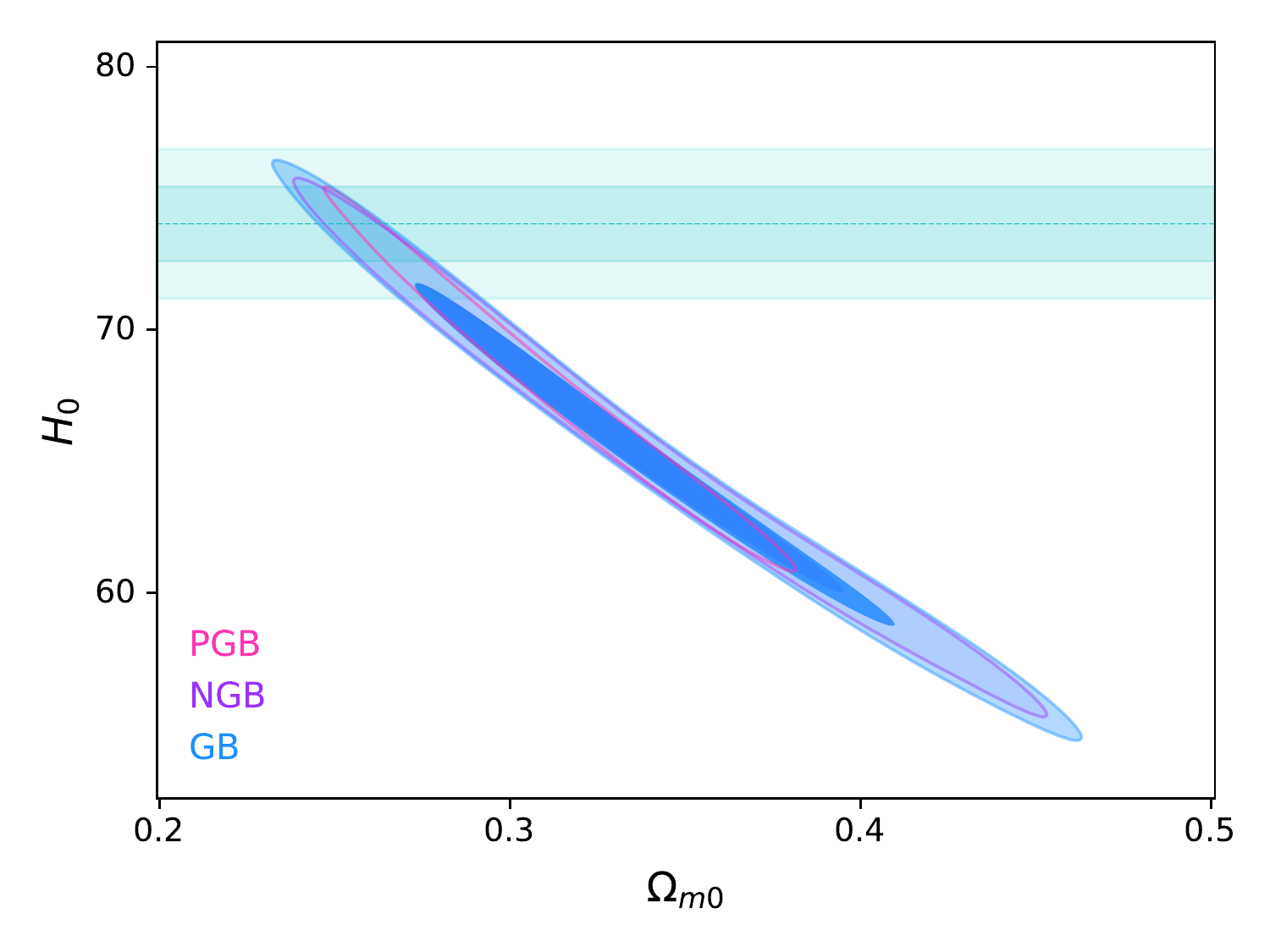}
	\caption{The $\Omega_{m0}$-$H_0$ parameter spaces constrained by the Planck-2018 CMB data are shown in the PGB (pink), NGB (purple) and GB (blue) models, respectively.  The horizontal cyan bands represents the direct measurement $H_0=74.03\pm1.42$ km s$^{-1}$ Mpc$^{-1}$ from the HST project \cite{Riess19}.}
	\label{f6}
\end{figure}

\section{The coincidence problem and effective equation of state}
It is interesting to study whether the coincidence problem can be resolved in the GB model. First of all, according to Eq.(\ref{4}), we derive the effective energy density of dark energy as follows
\begin{equation}
\rho_{de}=\frac{1}{8\pi G}(\Lambda-6\tilde{\alpha} H^4), \label{10}
\end{equation}
where the second term is from the contribution of Gauss-Bonnet invariant. Then, substituting Eq.(\ref{10}) into Eq.(\ref{5}), we obtain the effective pressure of dark energy 
\begin{equation}
p_{de}=\frac{1}{8\pi G}(8\tilde{\alpha} H^2\dot{H}+6\tilde{\alpha} H^4-\Lambda). \label{11}
\end{equation}
Furthermore, the density ratio between dark matter and dark energy is expressed as 
\begin{equation}
\frac{\rho_{cdm}}{\rho_{de}}=\frac{3H_0^2\Omega_{cdm0}(1+z)^3}{3H_0^2\Omega_{\Lambda0}-6\tilde{\alpha}H^4}. \label{12}
\end{equation}
It is easy to find that only when $\tilde{\alpha}<0$, can the densities of dark matter and dark matter reach the same order of magnitude earlier than those do in $\Lambda$CDM. The analysis result is shown in Fig.\ref{f1}. We find that when $\tilde{\alpha}=-10^{-5}$ dark matter shares the same density as dark energy in GB earlier than in $\Lambda$CDM. But if we choose a smaller $\tilde{\alpha}$ such as $-10^{-8}$, the density ratio of GB will behaves the same as that of $\Lambda$CDM. As a consequence, we conclude that, in theory, the GB model is able to partly resolve the coincidence problem. 

Furthermore, it is also interesting to study the effects of Gauss-Bonnet term on the effective EoS of dark energy. Combining Eq.(\ref{10}) with (\ref{11}), we have
\begin{equation}
\omega_{de}(z)=-1+\frac{8\tilde{\alpha}(1+z) H^2H^\prime}{6\tilde{\alpha} H^4-\Lambda}, \label{13}
\end{equation}
where the prime denotes the derivative with respect to $z$. From Eq.(\ref{6}), it is easy to see that $\tilde{\alpha}$ can be either positive or negative. Consequently, in GB, we should discuss the effective EoS of dark energy for the cases of $\tilde{\alpha}>0$ and $\tilde{\alpha}<0$, respectively. The corresponding numerical results are presented in Fig.\ref{f2}. We find that $\tilde{\alpha}$ can not be too large but very small when $\tilde{\alpha}>0$. Specifically, when $\tilde{\alpha}=10^{-3}$, to a large extent, the EoS deviates from $\Lambda$CDM. If $\tilde{\alpha}=10^{-8}$, it is always the same as $\Lambda$CDM at low redshifts and exhibits a clear deviation starting from $z=2$. If choosing a much smaller value $\tilde{\alpha}=10^{-16}$, the EoS tends to be same as $\Lambda$CDM. This provides a clue that $\tilde{\alpha}$ should be very small for us to constrain the GB model with observations. Similarly, for the case of $\tilde{\alpha}<0$, we also conclude that the absolute value of $\tilde{\alpha}$ should be very small in order to deviate small from $\Lambda$CDM. Very interestingly, we find that when $\tilde{\alpha}=-10^{-8}$, the EoS first becomes a phantom state starting from $z=0$ and then returns to $\Lambda$CDM at about $z=0.3$. If $\tilde{\alpha}$ is larger in the negative direction, e.g., $\tilde{\alpha}=-10^{-16}$, the EoS reduces to $\Lambda$CDM. This can also be viewed as a clue to implement cosmological constraints.

\section{Cosmological constraints}
In general, to construct a new modified gravity model, one should use the modified terms in the Lagrangian to replace the role of cosmological constant $\Lambda$. However, for the purpose of constructing a viable 4D GB gravity, the authors in Ref.\cite{Glavan:2019inb} include the small modified term $\alpha\mathcal{G}/(D-4)$ together with $\Lambda$ in the Lagrangian. Hence, we are very interested in whether the rescaling Gauss-Bonnet term can explain the cosmic acceleration alone. Base on this concern, Eq.(\ref{8}) with a vanishing $\Lambda$ is expressed as 
\begin{equation}
X(z)=1+4\tilde{\alpha}\left[\Omega_{m0}(1+z)^3+\Omega_{r0}(1+z)^4\right], \label{14}
\end{equation}  
and the relation Eq.(\ref{9}) reads as 
\begin{equation}
1+\tilde{\alpha}=\Omega_{m0}+\Omega_{r0}. \label{15}
\end{equation}
On the one hand, according to Eq.(\ref{7}), we have the condition $X(z)\geqslant0$. If we calculate $X(z)$ at the decoupling redshift $z_\star=1090.3$ \cite{Aghanim:2018eyx} and fix $\Omega_{m0}=0.3$ and $\Omega_{r0}=8.47\times10^{-5}$, then we obtain
\begin{equation}
\tilde{\alpha}\geqslant-4.91\times10^{-10}. \label{16}
\end{equation}
One the other hand, one can easily find that $\alpha\approx-0.7$ derived from Eq.(\ref{15}), which is strongly inconsistent with the inequality (\ref{16}). Therefore, from the background level only, we conclude that the rescaling Gauss-Bonnet term can not serves as dark energy alone. This explanation is the basic reason of the conclusion in Ref.\cite{Feng:2020duo} that the GB model with vanishing bare $\Lambda$ is roughly ruled out by the current observational limits of EoS of dark energy. 

In this study, we shall consider the case of nonvanishing $\Lambda$. Observing Eqs.(\ref{7}-\ref{9}), one can easily find that $\tilde{\alpha}$ should be a very small positive or negative value. Based on this concern, we consider three models to implement the Bayesian analysis, i.e., $\tilde{\alpha}>0$ (PGB), $\tilde{\alpha}<0$ (NGB) and free $\tilde{\alpha}$ (GB). Hereafter, we denote these three models as PGB, NGB and GB, respectively. Since we do not know the possible bounds of $\tilde{\alpha}$, the main purpose of constructing the PGB and NGB models 
is to determine the possible limitations of $\tilde{\alpha}$, in order to constrain $\tilde{\alpha}$ better in the GB model.
\begin{table*}[!t]
	\renewcommand\arraystretch{1.5}
	\caption{ The results of marginalized constraints on the cosmological parameters of PGB, NGB and GB models are shown, respectively. The symbols ``$\diamondsuit$'' denote the parameters that cannot be well constrained by observational datasets.    }
	\begin{tabular} { c |c| c |c| c|c |c }
		\hline
		\hline
		
		\multicolumn{2}{c|}{Parameter}                & $\Omega_{m0}$      & $\Omega_{b0}$          & $\alpha$            & $H_0$          &  $\sigma_8$                         \\
		\hline
		\multirow{3}{1cm}{PGB}  &     C  & $0.315\pm0.026$ & $0.0492\pm0.0041$  & $(0.20^{+1.00}_{-0.19})\times 10^{-16}$  &  $67.6^{+2.6}_{-3.2}$             &  $\diamond$                                                       \\
		\cline{2-7}
		&     R  & $0.275^{+0.066}_{-0.098}$  & $\diamond$   & $0.00030^{+0.00140}_{-0.00028}$ &    $70\pm20$ & $0.786^{+0.058}_{-0.070}$                                                        \\
		\cline{2-7}                        
		&     CBSHR  & $0.3008\pm0.0087$ & $0.0473\pm0.0012$  & $(0.08^{+0.68}_{-0.07})\times 10^{-16}$   &$68.88\pm0.90$  & $0.749\pm0.028$                                                   \\
		
		\hline
		\multirow{3}{1cm}{NGB}  &     C  & $0.334^{+0.034}_{-0.047}$  & $0.0522^{+0.0052}_{-0.0073}$ &  $(-2.5^{+2.4}_{-2.0})\times 10^{-21}$  & $65.8^{+4.2}_{-3.7}$  &$\diamond$                                                       \\
		\cline{2-7}
		&     R  & $0.27^{+0.11}_{-0.14}$   &   $\diamond$ & $-0.0070^{+0.0069}_{-0.0702}$ & $69\pm20$ &  $0.777\pm0.064$                                                       \\
		\cline{2-7}                       
		&     CBSHR  & $0.2997\pm0.0085$ & $0.0471\pm0.0012$ & $(-3.6^{+3.58}_{-5.40})\times 10^{-21}$ &  $68.99\pm0.89$ & $0.751\pm0.028$                                                           \\		                        
		\hline
		\multirow{3}{1cm}{GB}  &     C  & $0.340^{+0.041}_{-0.050}$  &  $0.0533^{+0.0064}_{-0.0079}$ &  $(4.2^{+3.3}_{-4.9})\times 10^{-17}$ & $65.2\pm4.3$ & $\diamond$                                                        \\
		\cline{2-7}
		&     R  &  $0.278^{+0.088}_{-0.092}$  &  $\diamond$  & $0.0004\pm0.0016$ & $68\pm21$ & $0.790\pm0.066$                                                     \\
		\cline{2-7}                        
		&     CBSHR   & $0.3004\pm0.0087$ & $0.0473\pm0.0012$ & $(1.2\pm5.2)\times 10^{-17}$ &$68.80\pm0.90$   &$0.750\pm0.028$                                                          \\

		\hline
		\hline
	\end{tabular}
	\label{t1}
\end{table*}

We will adopt the background and perturbation data to perform cosmological constraints on these models. Specifically, the background part consists of CMB, BAO and SNe Ia. Since the contribution of Gauss-Bonnet term is proportional $H^4$ in Eq.(\ref{4}), one can place a strong constraint on $\tilde{\alpha}$ with high redshift background data. Very interestingly,  the CMB data includes the distance information of comoving sound horizon at the decoupling redshift $z_\star$, and is able to give a tight restriction. Here we use the Planck-2018 distance prior from  TTTEEE$+$lowl$+$lowE$+$lensing data, i.e., compressed CMB data obtained in Ref.\cite{Zhai:2018vmm} to implement constraints. This dataset is denoted as ``C''. BAO as a standard cosmological ruler can constrain the expansion history of the universe after decoupling by measuring the position of the oscillations in the matter power spectrum at different redshifts, while breaking the degeneracies between parameters better. It is a clean signature and unaffected by other systematic uncertainties. Here we use the BOSS DR12 sample at three effective redshifts $z_{eff}=$ 0.38, 0.51 and 0.61 \cite{Alam:2016hwk}, the 6dFGS one at $z_{eff}=$ 0.106 \cite{Beutler:2011hx} and the SDSS-MGS one at $z_{eff}=$ 0.15 \cite{Ross:2014qpa}. This dataset is identified as ``B''. SNe Ia, the so-called standard candle, is a powerful tool to probe the expansion history of the universe. 
Up to now, the largest SNe Ia dataset is the ``Pantheon'' sample consisting of 1048 spectroscopically confirmed SNe Ia and covering the redshift range $z \in [0.01, 2.3]$. It is worth noting that under the assumption of a flat universe, this sample can be compressed into six $E^{-1}(z)$ points and is also able to constrain the background expansion well \cite{Riess:2017lxs}. For simplicity, we shall employ these compressed SNe Ia points in the numerical analysis and denote this dataset as ``S''. As a complementary background probe, we also include 31 cosmic chronometers \cite{Moresco:2016mzx}, which measure the Hubble parameter at different redshifts, in our constraint. We refer to this dateset as ``H''. 

For the first time, we would like to constrain these 4D Gauss-Bonnet models using the perturbation data, i.e., RSD, which includes the information of large scale structure. Here we adopt the so-called ``Gold-2018'' growth-rate dataset \cite{Basilakos:2016nyg} and denote this dataset as ``R''. To determine the value of $\tilde{\alpha}$ in a detailed way, we constrain the above three models using the dataset ``C'', ``R'' and a data combination of ``CBSHR'', respectively.

In order to perform the Bayesian analysis and obtain the posterior distributions of model parameters, we employ the Affine Invariant Markov chain Monte Carlo Ensemble sampler {\it EMCEE} \cite{ForemanMackey:2012ig} and analyze the chains with the package {\it GetDist} \cite{Lewis:2019xzd}.  

The results of marginalized constraints on the PGB, NGB and GB models are presented in Figs.\ref{f3}-\ref{f5} and Tab.\ref{t1}. By constraining the PGB and NGB models with Planck-2018 CMB data alone, we obtain a very tight bound $-4.5\times10^{-21}\leqslant\tilde{\alpha}\leqslant1.2\times10^{-16}$ at the $1\sigma$ confidence level, which exhibits a non-symmetric limitation by about five orders of magnitude in both the positive and negative directions. Enlarging this bound by six orders of magnitude as a prior for the GB model, we obtain the estimated value $\tilde{\alpha}=(4.2^{+3.3}_{-4.9})\times 10^{-17}$ for the case of free $\tilde{\alpha}$, while the non-symmetric property of $\tilde{\alpha}$'s error disappears. Very interestingly, we find that the $4.4\sigma$ $H_0$ tension between the directly local measurement from the Hubble Space Telescope (HST) \cite{Riess19} and indirectly global derivation from the Planck-2018 final release under $\Lambda$CDM \cite{Aghanim:2018eyx} can be greatly relieved to $2.17\sigma$, $1.86\sigma$ and $1.94\sigma$ level in the PGB, NGB and GB models, respectively (see also Fig.\ref{f6}).
We refer the readers to Ref.\cite{DiValentino:2021izs}, which is a complete review of $H_0$ tension. 
Subsequently, using the combined dataset CBSHR, we obtain a relatively conservative value $\tilde{\alpha}=(1.2\pm5.2)\times 10^{-17}$, which is consistent with $\Lambda$CDM within $1\sigma$ level. Note that this is, so far, the strongest constraint on the typical parameter $\tilde{\alpha}$. 
In natural units $h=c=k=1$, this constraint can be translated into $\alpha=(2.69\pm11.67)\times10^{48}$ eV$^{-2}$, which is tighter than the result obtained in Ref.\cite{Feng:2020duo} by at least one order of magnitude.  
From Fig.\ref{f5} and Tab.\ref{t1}, in the GB model, one can find that the constraint on $\tilde{\alpha}$ from RSD alone is much looser than those from CMB and CBSHR, and it gives a poor constraint on $H_0$. Furthermore, the constraints on the amplitude of matter clustering $\sigma_8$ are all compatible with Planck-2018 CMB final release at the $1\sigma$ confidence level \cite{Aghanim:2018eyx} but with about a ten times larger uncertainty. Meanwhile, we observe that the introduction of CBSH into R dataset leads to a smaller matter clustering effect in the large scale structure. 

\section{Large scale structure}
In the framework of conformal Newtonian gauge, when considering the scalar perturbations only, the perturbed FRW metric is expressed as \cite{Ma1995}
\begin{equation}
ds^2=a^2(\tau)\left[ -(1+2\Psi)d\tau^2+(1-2\Phi)\gamma_{ij}dx^idx^j  \right], \label{17}
\end{equation}
where $\Psi$ and $\Phi$ denote perturbed metric potentials and $\tau$ is the conformal time. The components of perturbed energy-momentum tensor are written as 
\begin{equation}
\delta T_0^0=-\tilde{\delta}\rho, \label{18}
\end{equation}
\begin{equation}
\delta T_0^i=-(1+c_s^2)\rho v^i, \label{19}
\end{equation}
\begin{equation}
\delta T_1^1=\delta T_2^2=\delta T_2^2=c_s^2\tilde{\delta}\rho, \label{20}
\end{equation}
where $\rho$ is the density of the fluid, $\tilde{\delta}=\delta\rho/\rho$ is the dimensionless density perturbation, $v$ is the velocity perturbation, and $c_s$ is the adiabatic sound speed of the fluid. 

For the non-relativistic matter component, its equation of state $\omega_m$ and squared adiabatic sound speed $c^2_{s(m)}$ both equal zero, namely $\omega_m=c^2_{s(m)}=\delta P_m/\delta \rho_m=0$. Since the energy conservation equation in the GB model is identical to that in GR, the perturbed conservation equation also remains the same as that in GR. Subsequently, for the matter component, the temporal and spatial components of perturbed Einstein field equation can be shown as 
\begin{equation}
\delta_m^\ast=3\Phi^{\ast}-\theta_m, \label{21}
\end{equation} 
\begin{equation}
\theta_m^\ast=k^2\Psi-\mathcal{H}\theta_m, \label{22}
\end{equation} 
where $\delta_m$ and $\theta_m$ are, respectively, the density and velocity perturbations of matter, $\mathcal{H}$ is the conformal Hubble parameter and the symbol ``$\ast$'' denotes the derivative with respect to the conformal time. The spatial off-diagonal component of perturbed field equation is given by
\begin{equation}
(2\tilde{\alpha}\mathcal{H}^2+a^2H_0^2)\Psi+(2\tilde{\alpha}\mathcal{H}^2-a^2H_0^2-4\tilde{\alpha}\mathcal{H}^\ast)\Phi=0. \label{23}
\end{equation}
One can easily find that this equation reduces to $\Psi=\Phi$ when $\tilde{\alpha}=0$. Substituting Eq.(\ref{23}) into Eqs.(\ref{21}) and (\ref{22}), in the sub-horizon limit $k\gg \mathcal{H}$ ($k$ is the comoving wave number), we obtain a second order differential equation for the density perturbation as 
\begin{equation}
\delta_m^{\prime\prime}+(\frac{H^\prime}{H}-\frac{1}{1+z})\delta_m^\prime+
\frac{3\Omega_{m0}(1+z)\left\{2\tilde{\alpha}E(z)\left[2(1+z)\frac{H^\prime}{H_0}-E(z)  \right]-1  \right\} }{2E^2(z)\left[2\tilde{\alpha}E^2(z)+1  \right]}\delta_m = 0 , \label{24}
\end{equation}
where the prime denotes the derivative with respect to the redshift $z$. 

\begin{figure}[htbp]
	\centering
	\includegraphics[scale=0.58]{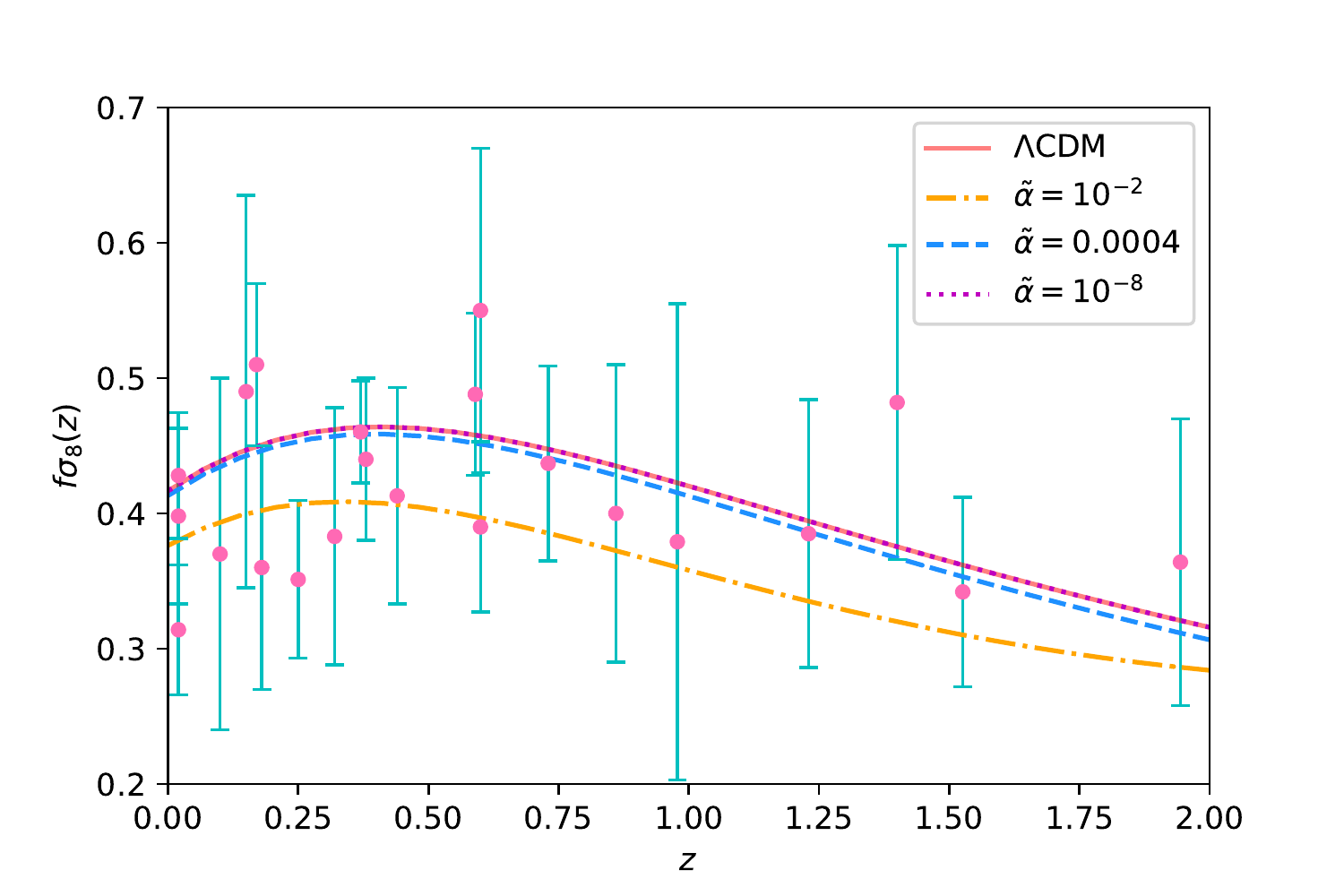}
	\includegraphics[scale=0.58]{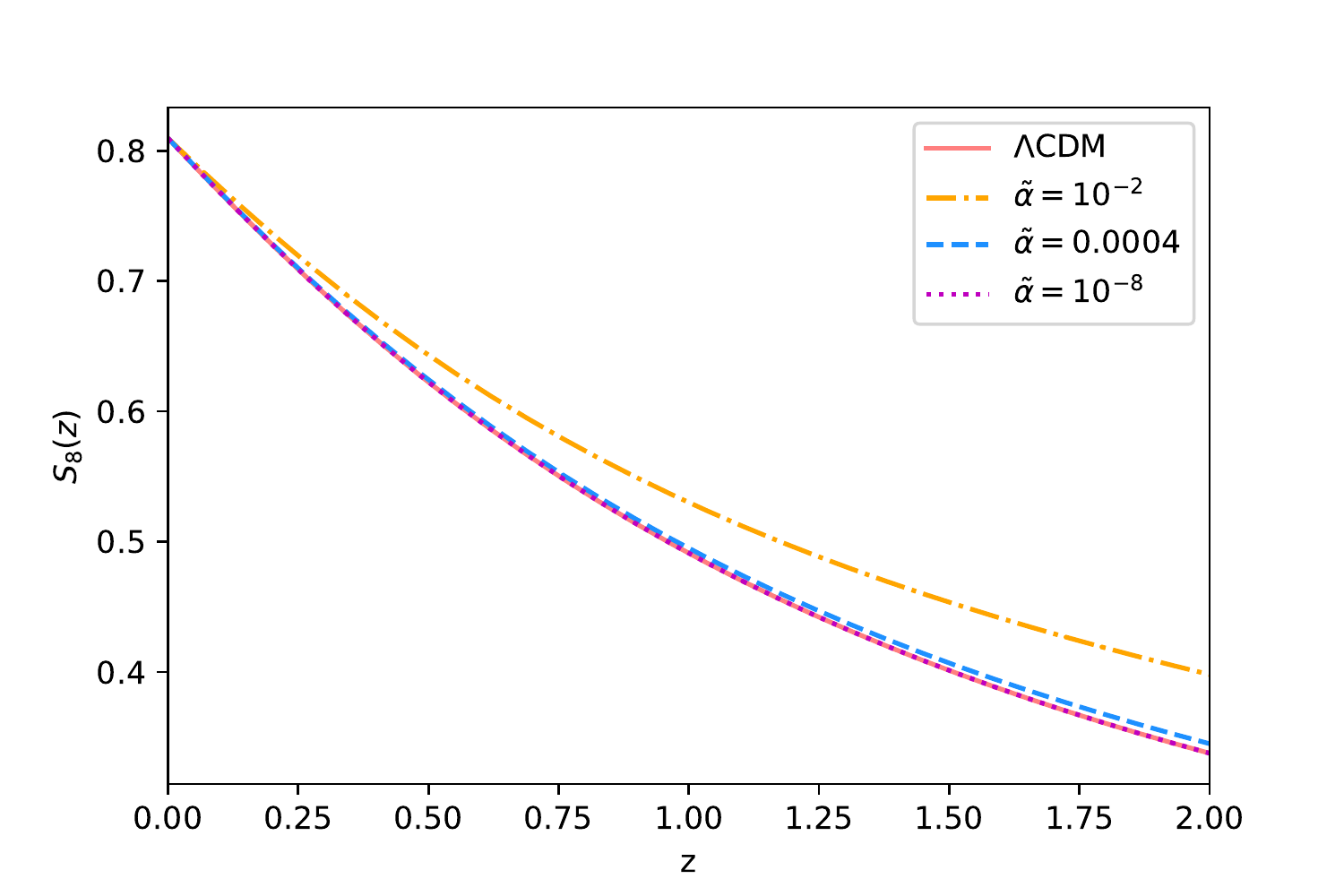}
	\caption{Two perturbation quantities $f\sigma_8(z)$ and $S_8(z)$ are shown as a function of redshift $z$, respectively. The solid (red) lines denote the $\Lambda$CDM model. The dash-dotted (orange), dashed (blue) and dotted lines are $\tilde{\alpha}=10^{-2}$, $0.0004$ and $10^{-8}$ in the GB model, respectively. The points with errors represent the RSD data. }
	\label{f7}
\end{figure}

To investigate the behaviors of large scale structure in the GB model, we define the following two perturbation quantities
\begin{equation}
f\sigma_8(z)=\sigma_8(z)\frac{\delta_m^\prime}{\delta_m} ,\label{25}
\end{equation}
\begin{equation}
S_8(z)= \sigma_8(z)\left(\frac{\Omega_{m0}}{0.3} \right)^{\frac{1}{2}} ,\label{26}
\end{equation}
where $\sigma_8(z)\equiv\sigma_8\delta_m(z)/\delta_m(0)$. By choosing appropriate initial conditions at the beginning of dark matter dominated epoch, we solve numerically Eq.(\ref{24}) and the corresponding results are shown in Fig.\ref{f7}. We find that a large value of $\tilde{\alpha}$ such as $10^{-2}$ in the GB model leads to a lower $f\sigma_8$ value at all redshifts than $\Lambda$CDM, can not explain the growth-rate dataset well, and gives a stronger effect of matter clustering at the same redshift than $\Lambda$CDM. The GB model with the constrained parameter $\tilde{\alpha}=0.0004$ by RSD data alone or a smaller value $10^{-8}$ almost exhibits the same evolutionary behaviors as $\Lambda$CDM. This implies that the constraining power of RSD data, which just can give restrictions on $\tilde{\alpha}\sim\mathcal{O}(-3)$ (see also Tab.\ref{t1}), is weak.    

\begin{figure}[htbp]
	\centering
	\includegraphics[scale=0.6]{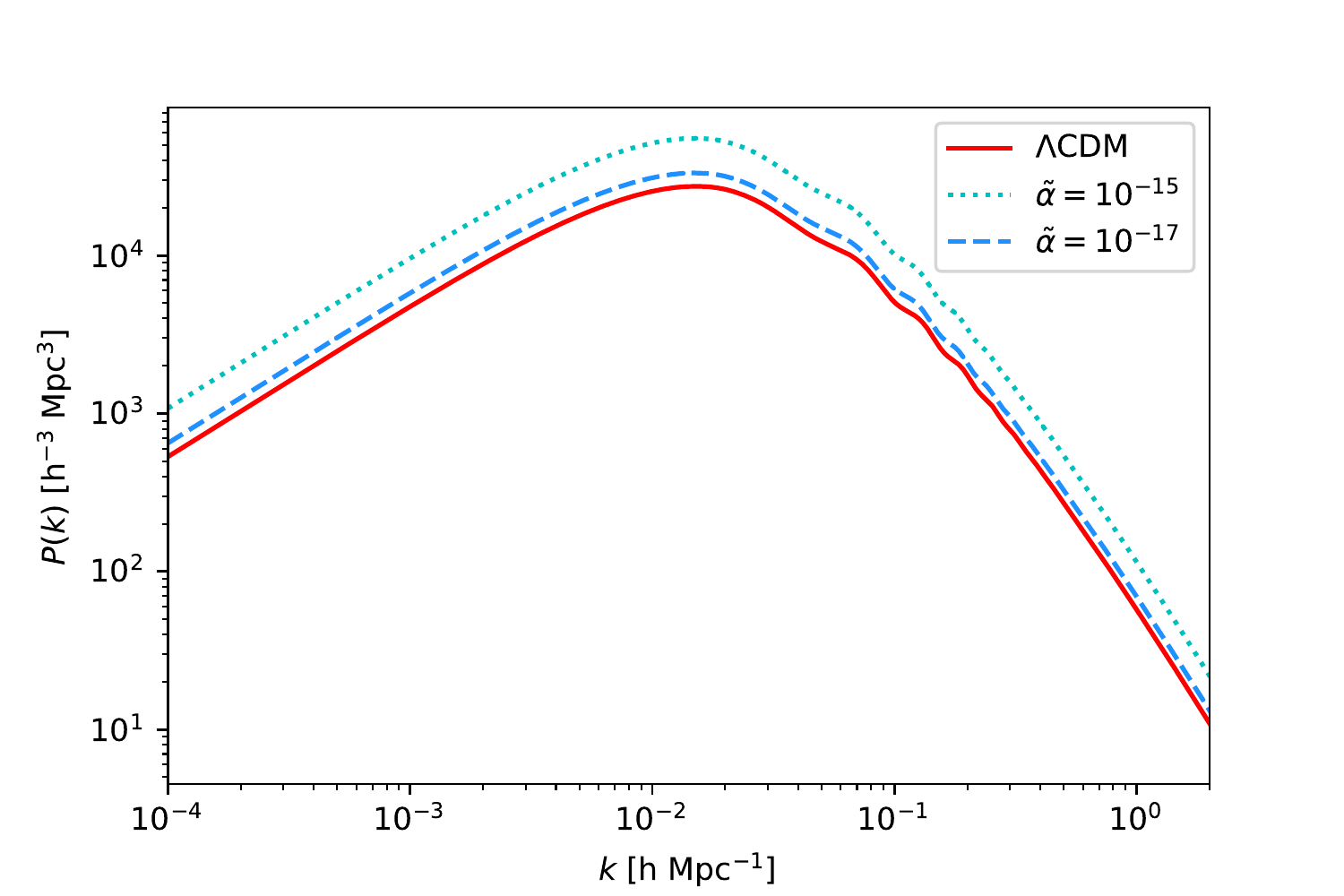}
	\caption{ The matter power spectra are shown for $\Lambda$CDM (solid), $\tilde{\alpha}=10^{-15}$ (dotted) and $10^{-17}$ (dashed), respectively.  }
	\label{f8}
\end{figure}

\begin{figure}[htbp]
	\centering
	\includegraphics[scale=0.58]{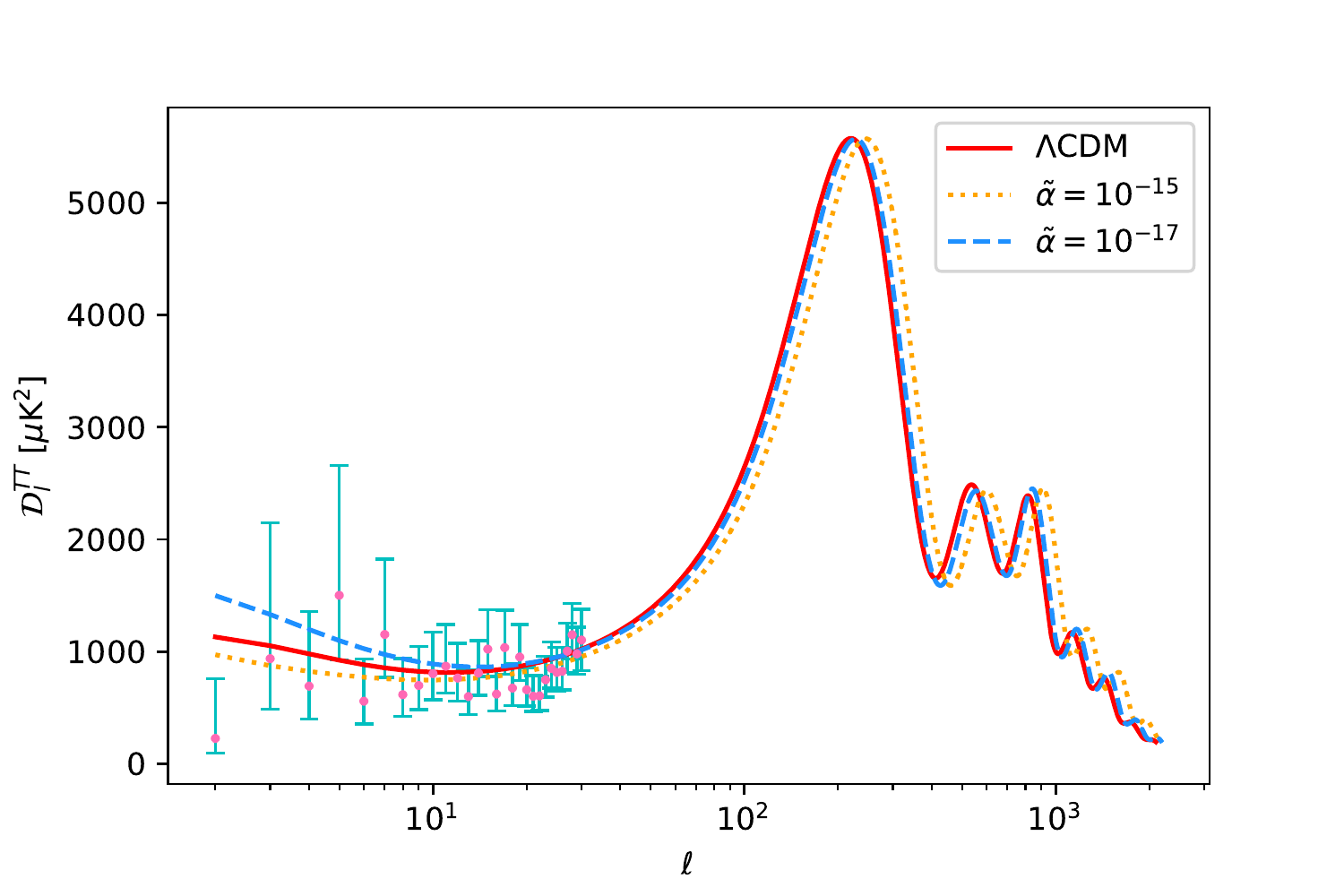}
	\includegraphics[scale=0.58]{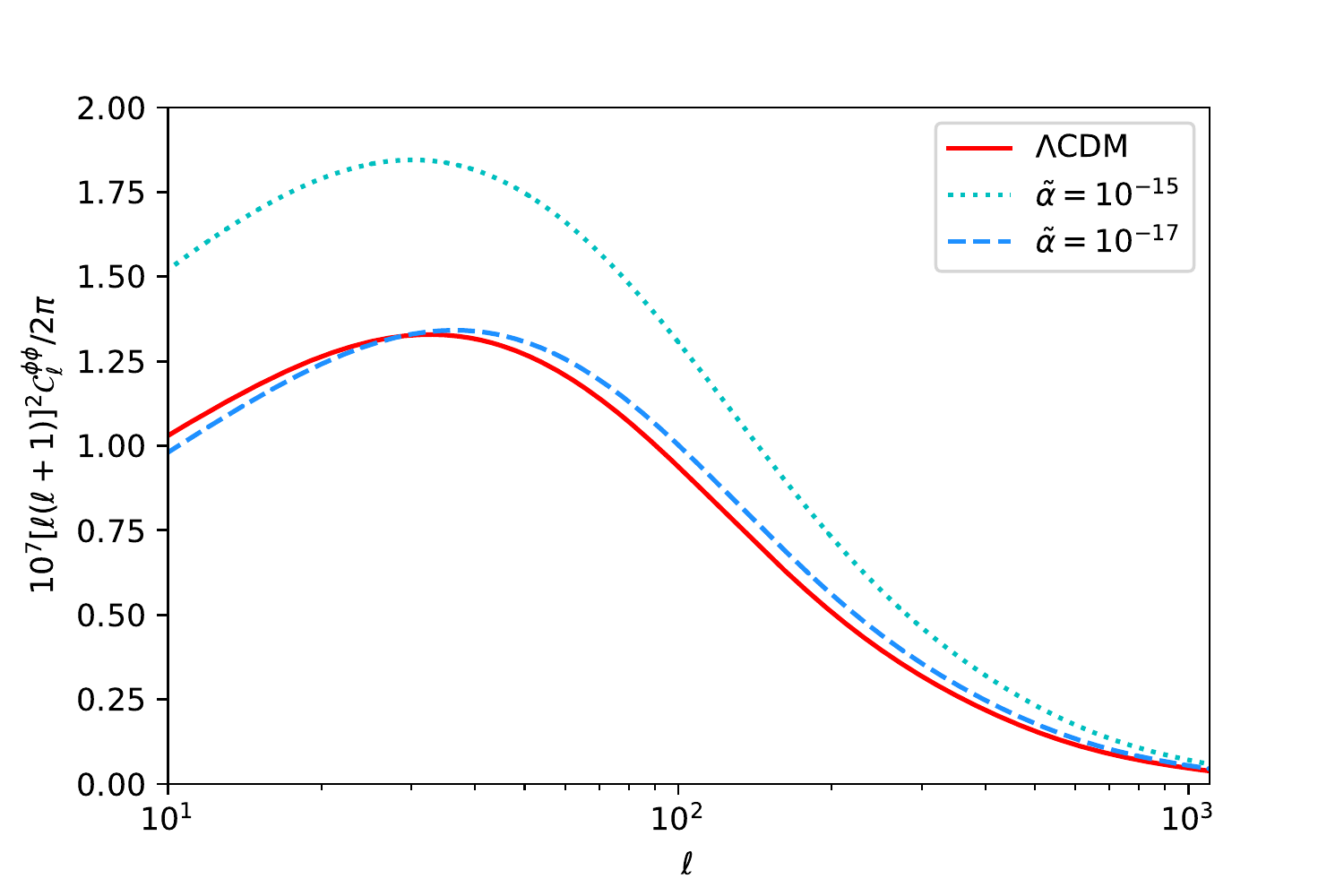}	
	\caption{ The CMB temperature power spectra and lensing potential power spectra ranging from multipole $\ell=2$ to 2200 are shown for $\Lambda$CDM (solid), $\tilde{\alpha}=10^{-15}$ (dotted) and $10^{-17}$ (dashed), respectively.  The points with errors represent the Planck-2018 temperature power spectrum data ranging from $\ell=2$ to 30.}
	\label{f9}
\end{figure}

To study further the matter power spectra, CMB temperature power spectra and CMB lensing potential power spectra, we modify the background and perturbation equations of the GB model in the public Boltzmann code CAMB \cite{Lewis2013} and the results are displayed in Figs.\ref{f8}-\ref{f9}. Note that we adopt the amplitude of primordial matter spectrum $A_s=2.1\times10^{-9}$, scalar spectral index $n_s=0.96$, pivot scale $k_{pivot}=0.05$ Mpc$^{-1}$, $\Omega_{b0}h^2=0.0226$, $\Omega_{cdm0}h^2=0.112$ and $H_0=70$ km s$^{-1}$ Mpc$^{-1}$ in CAMB. Here $h\equiv H_0/(100$ km s$^{-1}$ Mpc$^{-1})$. 

From Fig.\ref{f8}, we find a clear deviation from $\Lambda$CDM in the case of $\tilde{\alpha}=10^{-15}$, which has a larger power than $\Lambda$CDM  at all scales.  
In the left panel of Fig.\ref{f9}, we find that when $\tilde{\alpha}=10^{-15}$, although the CMB temperature power spectrum gives lower values than $\Lambda$CDM, it seems to be still compatible with the data points at large angular scales. However, the whole CMB temperature power spectrum has a obvious shift towards a larger multipole, which presents a clear deviation from $\Lambda$CDM. This indicates that one can rule out this value at small angular scales (see also Fig.1 in Ref.\cite{Aghanim:2018eyx}).
Furthermore, we also analyze the lensing potential power spectrum for the case of $\tilde{\alpha}=10^{-15}$, in the right panel of Fig.\ref{f9}, we find that this value has been ruled out by Planck-2018 lensing potential data (see also Fig.3 in Ref.\cite{Aghanim:2018eyx}). The above results are all consistent with our best constraint $(0.2\pm7.0)\times 10^{-17}$ and clearly rule out the GB model with $\tilde{\alpha}\geqslant10^{-15}$. It is worth noting that when $\tilde{\alpha}=10^{-17}$, the GB model always tends to be very close to $\Lambda$CDM in the analysis of matter, CMB temperature and lensing potential power spectra. In light of this, we need more high precision data to break the parameter degeneracy better, give tighter constraints, and then distinguish cosmological models in a more efficient way.     

\section{Discussions and conclusions}
A new Gauss-Bonnet gravity makes the Gauss-Bonnet invariant contribute nontrivially to gravitational dynamics in 4D spacetime by a regularization-like method. We are motivated by giving a correct and reasonable constraint on the free parameter $\alpha$ of the GB model in light of current cosmological observations. Using the joint constraint from cosmic microwave background, baryon acoustic oscillations, Type Ia supernovae, cosmic chronometers and redshift space distortions, we obtain the tightest constraint $\tilde{\alpha}=(1.2\pm5.2)\times 10^{-17}$, i.e., $\alpha=(2.69\pm11.67)\times10^{48}$ eV$^{-2}$ among various observational bounds from different information channels, which is tighter than previous limitation from the speed of gravitational wave by at least one order of magnitude. 

By studying the temperature and lensing potential power spectra of cosmic microwave background, we find that our constrained value of $\alpha$ is also supported by the Planck-2018 final data. However, due to large errors, the growth-rate data is unable to help improve the constraint.    

The improvement of our constraint mainly originates from the high precision distance data of cosmic microwave background at a high redshift, namely the decoupling redshift $z_\star=1090.3$ (see also Fig.\ref{f5} and Tab.\ref{t1}). Since the Gauss-Bonnet term $2\alpha H^4$ in Eq.(\ref{4}) must be small and the Hubble parameter $H$ becomes very large at $z_\star$, we obtain a very strong constraint. One can also find that combining other datasets with cosmic microwave background data just ameliorates the constraint a little.     

It is interesting that the serious $H_0$ tension between the local measurement from the Hubble Space Telescope (HST) \cite{Riess19} and global derivation from the Planck-2018 final data under the assumption of $\Lambda$CDM \cite{Aghanim:2018eyx} can be greatly alleviated from $4.4\sigma$  to $1.94\sigma$ level in the GB model.

In theory, the 4D GB gravity can partly resolve the coincidence problem. But if using the constrained small value of $\alpha$, the coincidence problem can not be resolved obviously. We also verify that the rescaling Gauss-Bonnet term is unable to serve as dark energy alone. It needs the help of the cosmological constant to explain the cosmic acceleration. Interestingly, one can also give a relatively large constraint by only using the behaviors of background evolution such as the effective equation of state of dark energy.  

In future, we expect that high precision data from the enhanced measurements of background expansion and large scale structure can help explore the nature of dark energy better.  

\section{Acknowledgements}
DW warmly thanks Yuan Sun and Changjun Gao for useful communications and discussions on gravitational theories. This work is supported by the Ministry of Science and Technology of China under Grant No.2017YFB0203300, National Nature Science Foundation of China under Grants No.11988101 and No.11851301. DFM thanks the Research Council of Norway for their support. Computations were performed on resources provided by UNINETT Sigma2 -- the National Infrastructure for High Performance Computing and Data Storage in Norway.

\end{document}